\documentclass[pra, twocolumn,showpacs,nofootinbibfloatfix,amsmath,amsfonts,amssymb]{revtex4-1}%
\usepackage{amsmath,amsfonts,amssymb,color}
\usepackage{amsthm}
\usepackage{leftidx}
\usepackage{graphicx}
\usepackage{xcolor}
\usepackage{dcolumn}
\usepackage{bm}
\usepackage{epstopdf}
\usepackage{epsfig}
\usepackage{physics}

\newcommand{\n}{\noindent}

\begin{document}
	\title{High-fidelity and long-distance entangled-state transfer with Floquet topological edge modes}
	\author{Senmao Tan}
	\affiliation{%
		Department of Physics, National University of Singapore, Singapore 117543
	}
	\author{Raditya Weda Bomantara}
	\email{phyrwb@nus.edu.sg}
	\affiliation{%
		Department of Physics, National University of Singapore, Singapore 117543
	}
	\author{Jiangbin Gong}%
	\email{phygj@nus.edu.sg}
	\affiliation{%
		Department of Physics, National University of Singapore, Singapore 117543
	}
	\date{\today}
	
	
	\vspace{2cm}
	
	\begin{abstract}
	
	We propose the generation of entangled qubits by utilizing the properties of edge states appearing at one end of a periodically driven (Floquet) superconducting qubit chain. Such qubits are naturally protected by the system's topology and their manipulation is possible through adiabatic control of the system parameters. By utilizing a Y-junction geometry, we then develop a protocol to perform high-fidelity transfer of entangled qubits from one end to another end of a qubit chain.  Our quantum state transfer protocol is found to be robust against disorder and imperfection in the system parameters. More importantly,  our proposed protocol also performs remarkably well at larger system sizes due to  nonvanishing gaps between the involved edge states and the bulk states, thus allowing us in principle to transfer entangled states over an arbitrarily large distance.    This work hence indicates that Floquet topological edge states are not only resourceful for implementing quantum gate operations, but also useful for high-fidelity and long-distance transfer of entangled states along solid-state qubit chains.

	\end{abstract}
	
	\maketitle
	
	\section{Introduction}
	
	Transferring quantum states from one place to another is an essential task in quantum information processing. The ubiquitous noise and device imperfections are however unavoidable and often limit the range for which a quantum state can be transferred with a good fidelity. Devising a scheme to effectively transfer quantum states over a long distance while minimizing the loss in fidelity has therefore been an active study since the last decade. Up to this date, various quantum state transfer (QST) protocols in many different platforms have been proposed, such as via strong coupling with photons \cite{qs1,qs3,qss1} or coherent transfer along a chain of qubits \cite{qs4,qs5,qs6,qs7,qs8,qs9,qs2}. In most cases, QST relies on the time-evolution of a specifically designed Hamiltonian, and as such perfect QST may require a very precise control over some of the system parameters, which may pose some difficulties in its large scale implementation.  Ref.~\cite{qs8} first proposed to use adiabatic control to facilitate robust quantum state transfer, but the dynamical phase induced by the adiabatic control field therein needs to be eliminated.
	
	In a seemingly separate area, topological phases of matter have emerged as a new paradigm for designing novel devices that are naturally immune to local disorder or imperfection. For example, topological insulators and superconductors possess the so-called edge states at their boundaries whose properties are insensitive to the specific details of the system \cite{Kit,TI,TI2,TI3}. Such inherent robustness of edge states makes them an ideal candidate for storing and processing quantum information.       Indeed, the use of edge states for quantum computing has been extensively studied and become an active research area on its own \cite{Kit,ising,Ivanov,tqc,tqc2,tqc3,RG,RG2,RG6}.  In addition, the time evolution of symmetry-protected topological edge states often induces trivial dynamical phases (e.g. zero dynamical phase), a feature that can often simplify protocol designs for quantum information processing.
	
	In recent years, several proposals to utilize edge states for QST have also emerged \cite{tqs,tqs2,tqs3}. In Ref.~\cite{tqs,tqs2}, chiral edge states of a topological material are used to transfer quantum information stored in one qubit to another distant qubit. The ability to control the coupling between the input and output qubits with the edge states is thus necessary in such proposals, which may not be straightforward to implement and may induce additional errors. An alternative approach to harnessing topological phenomena for QST would be to design a transfer protocol which directly controls the system in which the qubits are encoded. This was first explored in Ref.~\cite{tqs3}. There, logical qubits are encoded at the edges of a superconducting Xmon qubit chain with dimerized coupling. By adiabatically tuning the qubit-qubit couplings in a prescribed manner, a logical qubit located at one end of the chain can be transferred to the other end \cite{tqs3}. Owing to the topological nature of the edge states, both logical qubit encoding and QST aspects of the protocol are inherently robust against common perturbations or disorder \cite{tqs3}. It is thus natural to generalize such a proposal for transferring multiple edge-states-based qubits from one end to the other. This possibility has also been explored in Ref.~\cite{tqs3} by using trimerized instead of dimerized qubit-qubit couplings, although its implementation may be a challenge due to the extremely small energy gap in the protocol.
	
	In this paper, we present another approach for transferring entangled qubits which are encoded in the edge states of a periodically driven (Floquet) topological system. Our study is further motivated by the capability of Floquet topological phases to exhibit features with no static analogue, such as the existence of flat edge states at quasienergy (the analogue of energy in Floquet systems) $\pi/T$ \cite{RG,RG2,cref2,cref3,DG,RG3,RG4,RG5,LW,LW2,LW3} and anomalous edge states which do not satisfy the usual bulk-edge correspondence \cite{aes}. The former feature is particularly important in our present context, as the coexistence of flat (dispersionless) edge states at both quasienergy zero and $\pi/T$ naturally provides more channels for qubit encoding \cite{RG,RG2,RG6} and QST, as detailed below.
	
	Using the same Xmon qubit chain platform as that proposed in Ref.~\cite{tqs3}, periodic driving enables the existence of a pair of edge states at one end of the system, which allows the creation of entangled qubits. Adiabatic manipulation protocol of the qubit-qubit couplings can then be devised to transfer such entangled qubits from one end to the other while maintaining a very good fidelity due to its topological protection. The main difference between our approach and that of Ref.~\cite{tqs3} is twofold. First, entangled qubits can be prepared in a minimal setup with dimerized qubit-qubit couplings in our approach. Second, during the adiabatic manipulation, the edge states remain pinned at quasienergy zero and $\pi/T$. As a result,  large quasienergy gaps between the logical qubits and the rest of the qubits are maintained throughout the protocol, which is necessary for adiabaticity to hold. Our proposal thus demonstrates that while Floquet topological phases enable more qubits to be encoded as compared with their static counterpart under the same physical constraints, such qubits can also be transferred from one place to another using an approach similar to that in typical static systems.
	
	This paper is structured as follows. In Sec.~\ref{model}, we introduce the model studied in this paper, briefly review the emergence of zero and $\pi$ modes in the model and the topological invariants that characterize them, and set up some notation. In Sec.~\ref{state prep}, we propose a protocol to generate entangled qubits from the ground states of the underlying static model by utilizing the properties of the zero and $\pi$ modes that exist after the periodic driving is turned on. In Sec.~\ref{QST}, we present a protocol to transfer an entangled qubit from one end to another along a Y-shaped chain of periodically driven Xmon qubits. In Sec.~\ref{disc}, we verify the robustness of our QST protocol in the presence of disorder and imperfection, then we also show how our proposal can be adapted to improve previous QST protocol in Ref.~\cite{tqs3}, enabling high fidelity QST even for long qubit chains. Finally, we conclude our paper and present potential future directions in Sec.~\ref{conc}.
	
	\section{Model and qubit encoding}
	\label{model}
	
	\begin{figure}[ht]
	\centering
	\includegraphics[angle =0,width=0.45\textwidth]{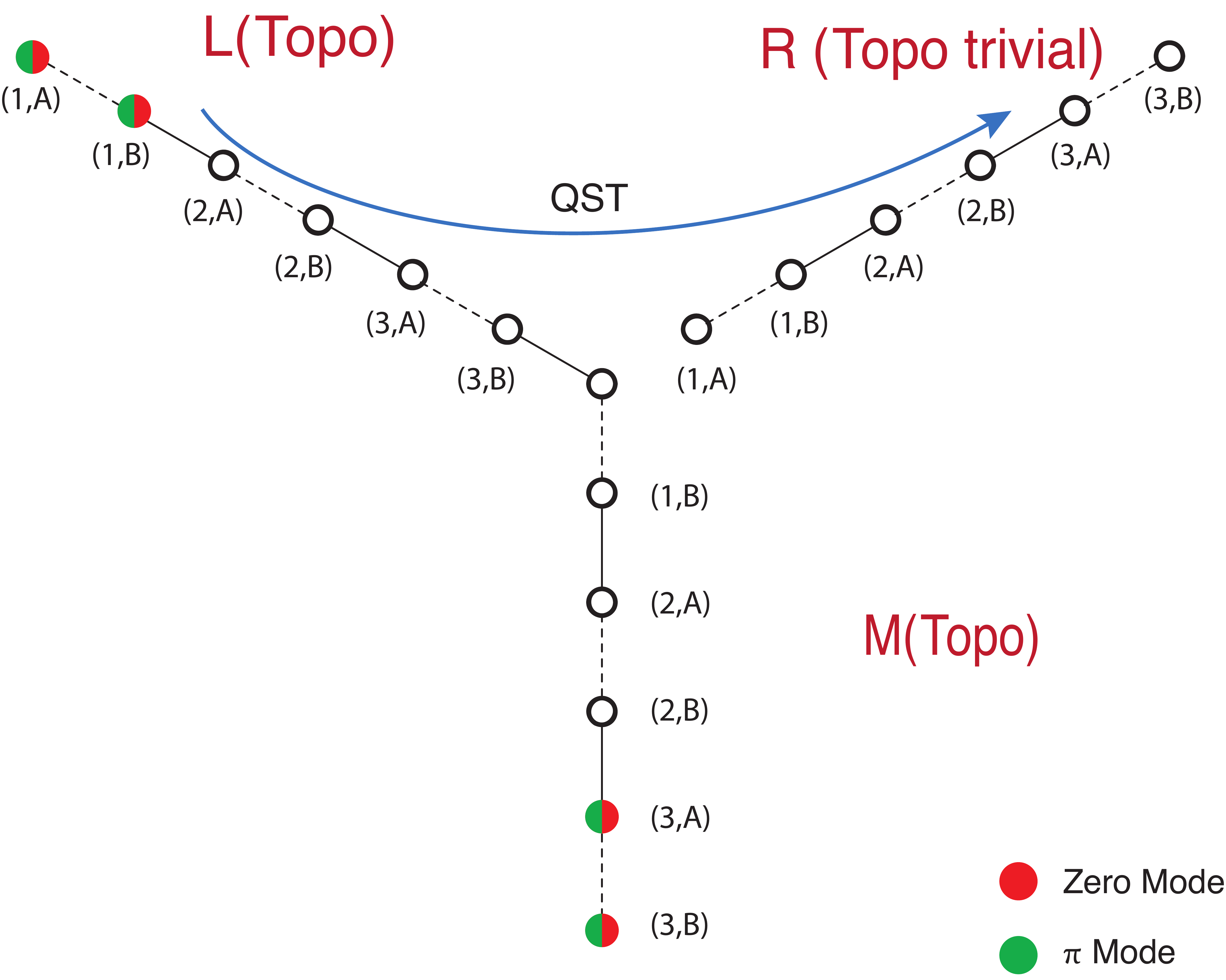}
	\caption{A chain of superconducting qubits with dimerized couplings arranged in a Y-junction geometry. In a certain regime of system parameters, a pair of zero and $\pi$ edge modes may emerge at one end of the L and M branches, which in \emph{the ideal case} appear as antisymmetric and symmetric superpositions of the first (or the last) two qubits respectively (see Eq.~(\ref{zpi})).}
	\label{fig:Y junction}
	\end{figure}
	We consider a chain of superconducting qubits arranged in a $Y$-junction geometry with dimerized nearest-neighbor time-periodic couplings, as depicted in Figure \ref{fig:Y junction}. Each circle represents a superconducting Xmon qubit and is categorized into sublattice A or B based on how it couples with its two neighbouring qubits. The dashed and solid lines mark two different coupling strengths between two such qubits, which are referred to as intra- and inter-lattice coupling respectively. L, R and M label the three branches of the system, each of which is described by a Hamitonian of the following form,
	\[ H^{(c)}(t)  = \begin{cases}
	H_1^{(c)}(t) & \text{for}\ (m-1)T<t<(m - 1/2)T, \\
	H_2^{(c)}(t) & \text{for}\ (m-1/2)T<t<m T.
	\end{cases}
	\]
	\begin{align}
	H_1^{(c)}(t) &= \sum_{j}\left(-J^{c}_{\mathrm{intra},j}\sigma^{c\dagger}_{B,j} \sigma^{c}_{A,j} -   J^{c}_{\mathrm{inter},j} \sigma^{c\dagger}_{B,j} \sigma^{c}_{A,j+1}  + h.c.\right),\\
	H_2^{(c)}(t) &= \sum_{j}\left(-j^{c}_{\mathrm{intra},j}\sigma^{c\dagger}_{B,j} \sigma^{c}_{A,j} -   j^{c}_{\mathrm{inter},j} \sigma^{c\dagger}_{B,j} \sigma^{c}_{A,j+1}  + h.c.\right), \label{ham}
	\end{align}
	where $c\in\{L,R,M\}$ labels one of the three branches, A and B are the indices of the sublattice site, $\sigma^{\dagger}_{S,j} = \ket{e}\bra{g}_{S,j}$ is the qubit raising operator at sublattice $S$ of unit cell $j$, $|g\rangle_j$ and $|e\rangle_j$ are the ground and excited states of the $j$-th Xmon qubit, $J_{\mathrm{intra}}$ ($J_{\mathrm{inter}}$)  and $j_{\mathrm{intra}}$ ($j_{\mathrm{inter}}$) are the intra-lattice (inter-lattice) coupling strength in $H_1$ and $H_2$ respectively. Unless otherwise specified, we take
	$J^{c}_{\mathrm{intra},j} = J_1^{c}=J_1, J^{c}_{\mathrm{inter},j} = J_2^{c}=J_2, j^{c}_{\mathrm{intra},j} = j_1^{c}=j_1$, and $j^{c}_{\mathrm{inter},j} = j_2^{c}=j_2$ in the following.
	
	The spectral properties of such a time-periodic system are characterized by quasienergies, defined as the eigenphase of the one-period propagator (Floquet operator \cite{Flo1,Flo2}),
	\begin{eqnarray}
	U \ket{\epsilon} &=& \exp(-i\epsilon T) \ket{\epsilon}, \nonumber  \\
	U &\equiv & \mathcal{T} \exp(\int_{t_0}^{t_0+T} -\frac{\mathrm{i}H(t)}{\hbar} dt), \label{flo}
	\end{eqnarray}
	where $\mathcal{T}$ is the time-ordering operator, $\epsilon$ is the quasienergy and $\ket{\epsilon}$ is the Floquet eigenstate with quasienergy $\epsilon$. By construction, $\epsilon$ is only defined modulo $\frac{2\pi}{T}$. As such, edge states may form not only in the gap around quasienergy zero which are commonly found in static systems,  but also in the gap around quasienergy $\pi/T$.
	
\begin{figure}[ht]
		\centering
		\includegraphics[angle =0,width=0.5\textwidth]{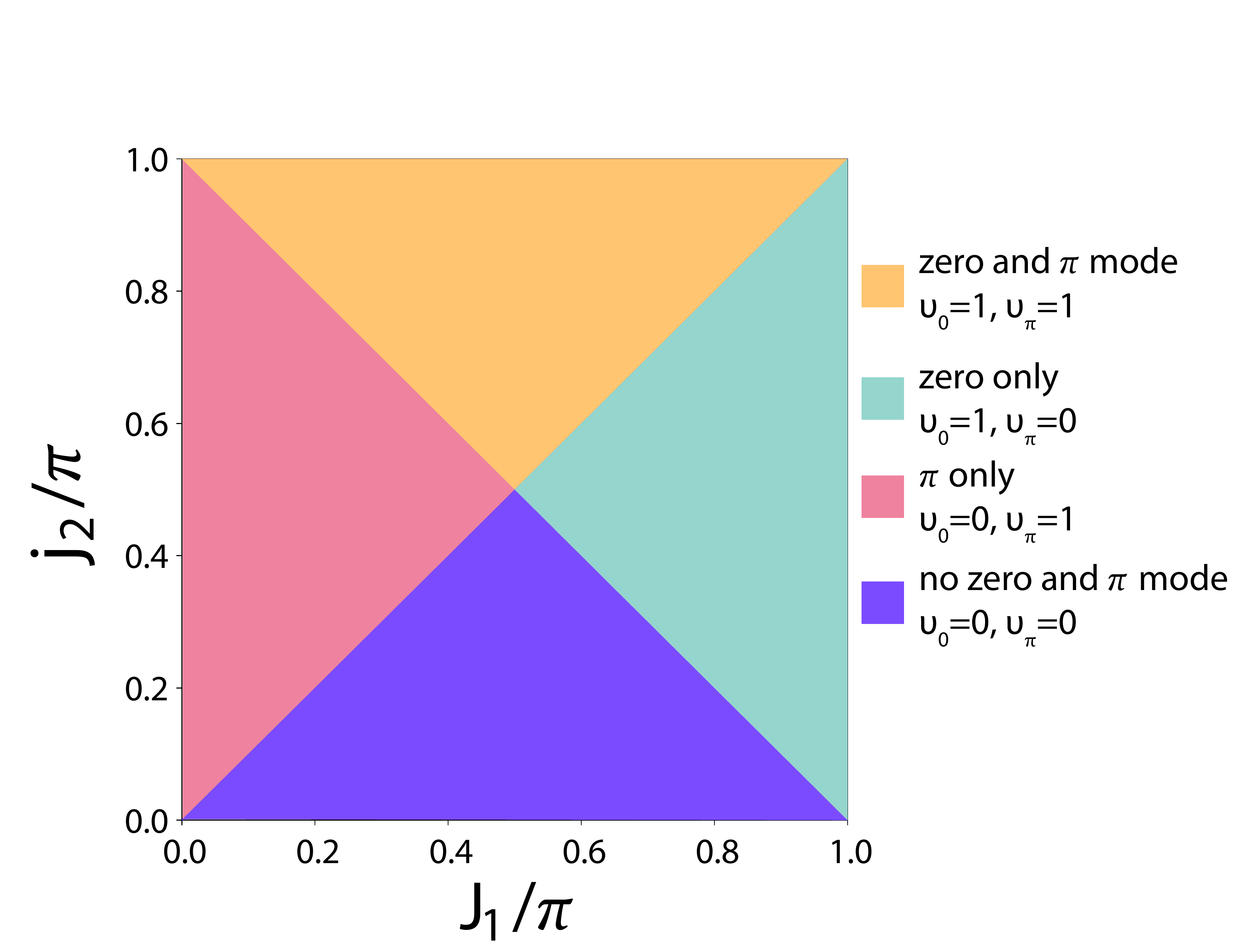}
		\caption{The phase diagram of the topological invariants $\upsilon_{0}$ and $\upsilon_{\pi}$ as a function of $J_1$ and $j_2$, where $J_2 = j_1 = 0$.}
		\label{fig:number of zero mode}
	\end{figure}

	In the momentum space, the two Hamiltonians defined in Eq.~(\ref{ham}) take the form
	\begin{equation}
	\mathcal{H}_\alpha^{(c)}(k,t) = -(h_a + h_b\cos(k))\tau_x +h_b\sin(k) \tau_y \;, \label{ham2}
	\end{equation}
	where $\alpha=1,2$ and $\tau$'s are Pauli matrices acting in the sublattice space. It then follows that our system possesses a chiral symmetry defined by the operator $\Gamma=\tau_z$, i.e., $\left\lbrace\mathcal{H}_\alpha^{(c)}(k,t),\tau_z\right\rbrace=0$, which pins its edge states at exactly zero and/or $\pi/T$ quasienergies (termed zero and $\pi$ modes respectively), whose numbers are determined by the topological invariants defined in \cite{cref2,cref3}. We can calculate these topological invariants by first writing down our momentum space Floquet operator $U(k)$ in the symmetric time frame (accomplished by taking $t_0=T/2$ in Eq.~(\ref{flo})),
	\begin{align}
	\label{sym time}
	U(k) &= F(k)G(k),\nonumber\\
	F(k) &= \exp(-i H_2(k)/2) \times \exp(-i H_1(k)/2),\nonumber\\
	G(k) &= \exp(-i H_1(k)/2) \times \exp(-i H_2(k)/2),
	\end{align}
	such that $F(k)=\Gamma^\dagger G(k)^\dagger \Gamma$ (we take $\hbar=\frac{T}{2}=1$ from here onwards). $F$ can then be represented by a $2\times2$ matrix in the canonical ($\Gamma=\tau_z$) basis as
	\begin{align}
	F(k) =
	\begin{pmatrix}
		a(k) & b(k) \\
		c(k) & d(k)
	\end{pmatrix},
	\end{align}
	and the topological invariants
	\begin{align}
	&\upsilon_{0} = \frac{1}{2\pi i } \int_{-\pi}^{\pi} dk\left(b^{-1}\frac{d}{dk}b\right),\\
	&\upsilon_{\pi} = \frac{1}{2\pi i } \int_{-\pi}^{\pi} dk\left(d^{-1}\frac{d}{dk}d\right), \label{zpi}
	\end{align}
	directly count the number of quasienergy zero and $\pi$ modes respectively. In Fig.~\ref{fig:number of zero mode}, we numerically compute $\upsilon_{0}$ and $\upsilon_{\pi}$ under some representative parameter values, which we have also analytically verified in Appendix~\ref{zpical}.
	
	In the following, we will mostly be interested in the yellow and purple regime of Fig.~\ref{fig:number of zero mode}, corresponding to the presence and absence of both zero and $\pi$ modes respectively. In particular, we initialize our system such that branches L and M are in the yellow regime, whereas branch R is in the purple regime. For analytical solvability, we will further consider the following parameter values (referred to as \emph{the ideal case}) for the yellow regime: $J_1 = i\pi/2, j_2 = i \pi, J_2 = j_1 = 0$ and for the purple regime: $J_1 = i \pi / 2, j_2 = J_2 = j_1 = 0$, in the presentation of our state preparation and QST protocols, where such imaginary couplings are indeed realizable in superconducting qubit setups \cite{qs2}. We will however show, through some numerical calculations, that such fine tuning is not necessary in the actual implementation of our protocols.
	
	In the ideal case, one pair of zero and $\pi$ modes is localized at the first two qubits (the first unit cell) of the L branch,
	\begin{align}
	\ket{0}^{(L)} &= \left(\ket{eg}_{1}^{(L)} - \ket{ge}_{1}^{(L)} \right)\otimes G^{\prime},\nonumber\\
	\ket{\pi}^{(L)} &= \left(\ket{eg}_{1}^{(L)} + \ket{ge}_{1}^{(L)} \right)\otimes G^{\prime}, \label{zpi2}
	\end{align}
	where subscript is the site index and $G^{\prime}=\prod_{i\ne 1} \ket{gg}_i$ denotes the ground state of the other Xmon qubits in the system. For brevity, $G^{\prime}$ is suppressed in the rest of this paper. In a similar fashion, note that there is also another pair of zero and $\pi$ modes in the M branch localized at the last two qubits (see Fig.~\ref{fig:Y junction}), which we will not discuss further in what follows. Finally, we note that the zero and $\pi$ edge modes defined above already represent two maximally entangled states between two qubits. Consequently, the task of preparing an entangled state then reduces to the task of preparing a Floquet edge state, the latter of which can be accomplished via a protocol introduced in the next section.

	\section{Entangled qubits generation}
	\label{state prep}
	
	\begin{figure}[ht]
		\centering
		\includegraphics[angle =0,width=0.5 \textwidth]{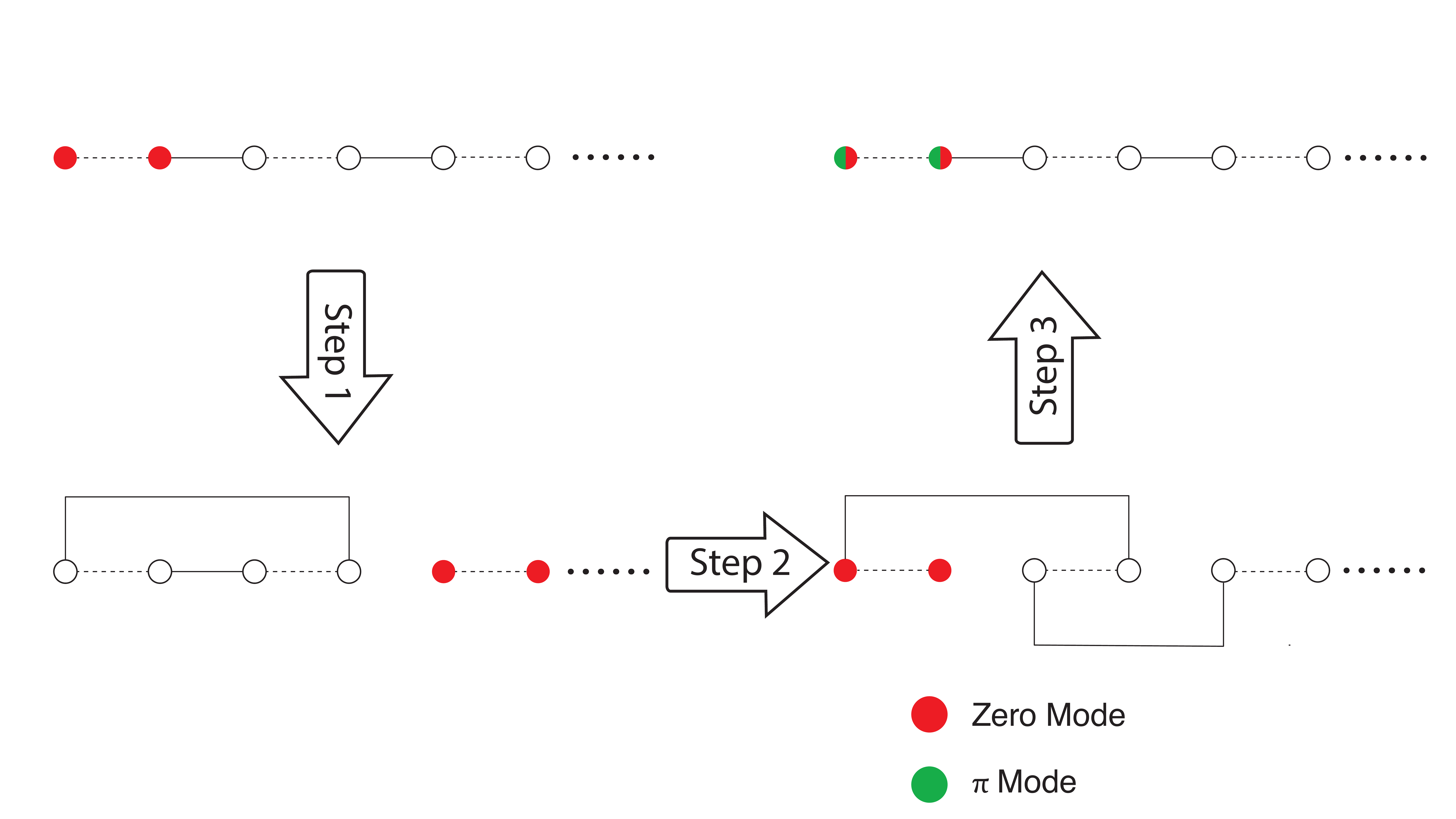}
		\caption{Schematic of the adiabatic protocol resulting in a $\pi/4$ rotation in the subspace spanned by the zero and $\pi$ modes. Here we focus on the L branch and highlight only the evolution of zero mode (red filled circles), which at the end transforms into a superposition of zero and $\pi$ modes (red and green half-filled circles). Dashed (solid) lines represent the qubit-qubit coupling appearing in $H_1$ ($H_2$). }
		\label{fig:braiding}
	\end{figure}
	Suppose that our system starts in a ground state of a static Hamiltonian $H=H_2^{(L)}+H_2^{(M)}+H_1^{(R)}$, before we switch on the periodic driving at some time $t_0$. Without loss of generality, we may assume that this corresponds to the initial state of $|\psi_0\rangle = |eg\rangle_1^{(L)}\propto |0\rangle^{(L)}+ |\pi\rangle^{(L)}$, which is thus a simple product state. Our objective in this section is to devise a protocol based on a series of adiabatic variations of some qubit-qubit couplings in the system, such that the above initial state evolves to either $|0\rangle^{(L)}$ or $|\pi\rangle^{(L)}$ at the end of the protocol. This can be accomplished by adapting the protocol introduced by two of us in Ref.~\cite{RG} which amounts to transforming zero and $\pi$ modes as $\ket{0}^{(L)}\rightarrow (\ket{0}^{(L)} - \ket{\pi}^{(L)})/\sqrt{2}$ and  $\ket{\pi}^{(L)}\rightarrow (\ket{0}^{(L)} + \ket{\pi}^{(L)})/\sqrt{2}$. To this end, it suffices to restrict our attention to branch L, so that we will remove the $(L)$ index in the following. We will now present our protocol in three steps, which is also summarized in Fig.~\ref{fig:braiding}.
	
	\textit{In step 1}, we adiabatically deform the Hamiltonian stroboscopically (by slowly varying it at every period) to move the zero and $\pi$ modes from the first to the third unit cell in the $L$ branch. This is accomplished by setting $j_{\mathrm{inter},2} = j_2 \cos\phi$ and at the same time introducing a new coupling $h_2^{(1)} = j_2\sin\phi\ \sigma^{\dagger}_{A,1} \sigma_{B,2} + h.c.$ into $H_2$, where $\phi$ is the adiabatic parameter which is swept from 0 to $\pi/2$. It can be shown that the zero and $\pi$ modes at any stroboscopic time take the form \cite{RG},
	\begin{align}
	\ket{0} &= \cos \phi \left(\ket{eg}_{1} - \ket{ge}_{1} \right) - \sin \phi \left(\ket{eg}_{3} - \ket{ge}_{3} \right),\label{zero_braid}\\
	\ket{\pi} &= \cos \phi \left(\ket{eg}_{1} + \ket{ge}_{1} \right) - \sin \phi \left(\ket{eg}_{3} + \ket{ge}_{3} \right).
	\end{align}
	At the end of this step, $\ket{0}$ adiabatically changes from $(\ket{eg}_{1} - \ket{ge}_{1}) $  to  $\left(-\ket{eg}_{3} + \ket{ge}_{3} \right)$, whereas $\ket{\pi}$ transforms from $(\ket{eg}_{1} + \ket{ge}_{1})$  to  $-\left(\ket{eg}_{3} + \ket{ge}_{3} \right)$, i.e., both zero and $\pi$ modes are now shifted to the third unit cell as intended. \\
	
	\textit{In step 2}, starting from the end of step 1, we continue to adiabatically deform the system's Hamiltonian by taking $j_{\mathrm{inter},1} = j_2 \cos \phi$ and introducing a new term $h_2^{(2)} = j_2 \sin \phi\ \sigma^{\dagger}_{A,2} \sigma_{A,3} + h.c.$ into $H_2$, where $\phi$ again changes slowly every period from $0$ to $\pi/2$ at the end of this step. We can again show that at any stroboscopic time \cite{RG},
	\begin{align}
	\ket{0} &= -\sin \phi \left(\ket{eg}_{1} + \ket{ge}_{1} \right) + \cos \phi \left(-\ket{eg}_{3} + \ket{ge}_{3} \right),\\
	\ket{\pi} &= \sin \phi \left(-\ket{eg}_{1} + \ket{ge}_{1} \right) - \cos \phi \left(\ket{eg}_{3} + \ket{ge}_{3} \right).
	\end{align}
	That is, $\ket{0}$ transforms to $-\ket{eg}_{1} - \ket{ge}_{1} $, whereas $\ket{\pi}$ transforms to $\left(-\ket{eg}_{1} + \ket{ge}_{1} \right)$ at the end of this step.
	\\
	
	\textit{In step 3}, we recover the system's original Hamiltonian by returning $j_{\mathrm{inter},1}$ and $j_{\mathrm{inter},2}$ back to their original values as $j_{\mathrm{inter},1} = j_{\mathrm{inter},2} = j_2 \cos \phi$ and slowly decreasing $h_2^{(1)}$ and $h_2^{(2)}$ to $0$ as $h_2^{(1)} \rightarrow h_2^{(1)} \cos^2\phi$ and  $h_2^{(2)} \rightarrow h_2^{(2)} \cos^2\phi$ with $\phi$ being slowly swept from $0$ to $\pi/2$. However, in order to induce a nontrivial rotation in the subspace spanned by zero and $\pi$ modes, we only tune the adiabatic parameter $\phi$ \emph{every other period}. As demonstrated in Ref.~\cite{RG}, this leads to the transformation $\ket{0}\rightarrow (\ket{0} - \ket{\pi})/\sqrt{2}$ and  $\ket{\pi}\rightarrow (\ket{0} + \ket{\pi})/\sqrt{2}$ at the end of this step, thus completing our protocol.\\
	\begin{figure}[ht]
		\centering
		\includegraphics[angle =0,width=0.5 \textwidth]{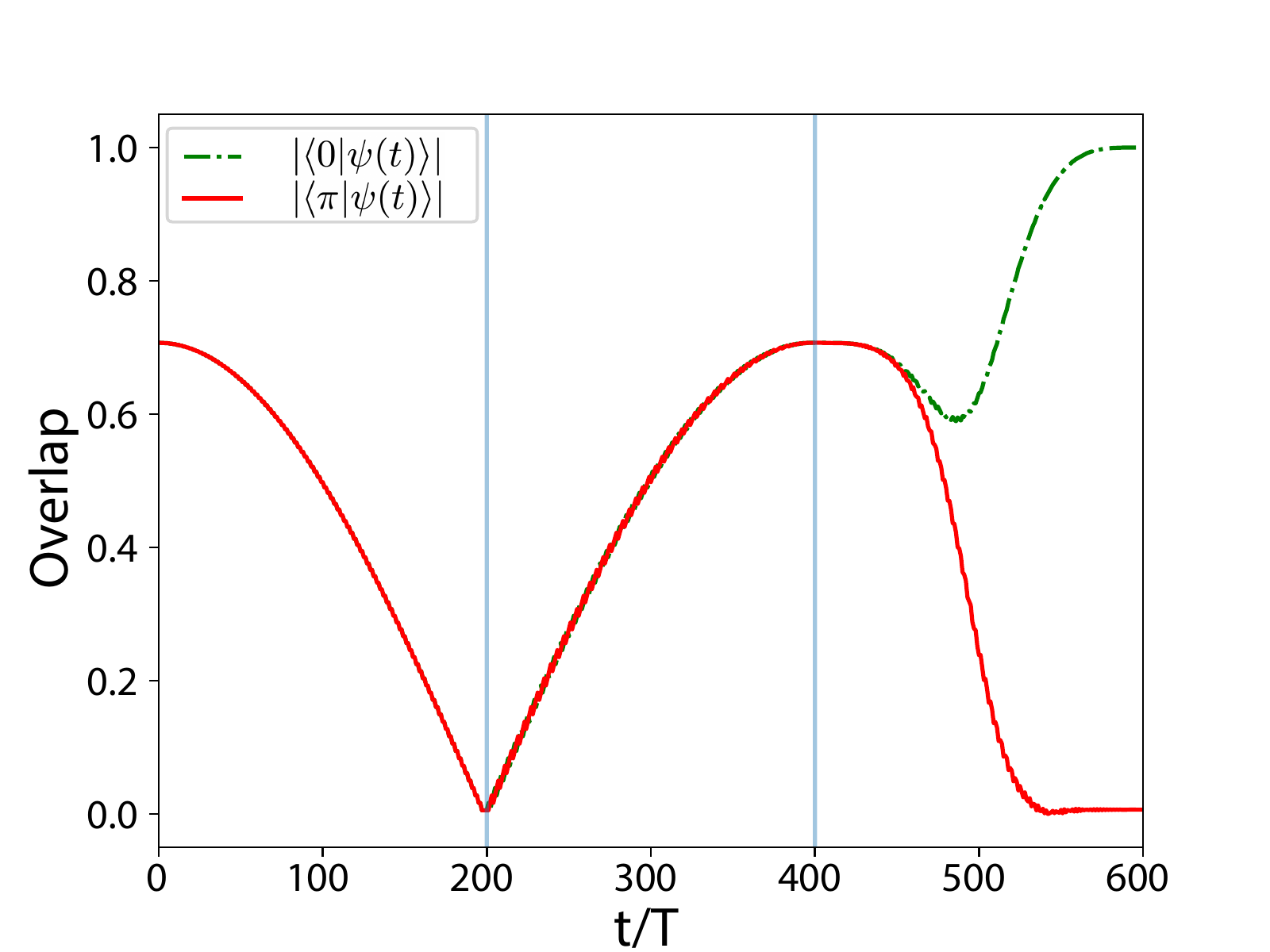}
		\caption{Time evolution of the overlap $|\langle 0|\psi(t)\rangle |$ and $|\langle \pi|\psi(t)\rangle |$ for a state initially prepared as a superposition of zero mode and $\pi$ mode, i.e., $|\psi(0)\rangle = \frac{1}{\sqrt{2}} \left(\ket{0} + \ket{\pi}\right)$. Each step takes 200 periods to complete.}
		\label{fig:overlap}
	\end{figure}
	
	
To explicitly demonstrate how an entangled state can be generated via the protocol above, in Fig.~\ref{fig:overlap} we numerically plot the overlap between the state $|\psi(t)\rangle$, initially prepared in the product state $|eg\rangle_1$, and the zero ($\pi$) mode of the original system, both of which represent a maximally entangled two qubit states (see Eq.~(\ref{zpi2})).  In particular, since its overlap with the zero mode becomes unity at the end of the adiabatic protocol, our state transforms into $|0\rangle =|eg\rangle_1 - |ge\rangle_1$. To generate another entangled state $|\pi\rangle = |eg\rangle_1+|ge\rangle_1$, we can simply perform exactly the same protocol two more times.

In principle, we can also prepare a more generic entangled state by exploiting the dynamical evolution of the Floquet operator. In particular, since $|0\rangle$ and $|\pi\rangle$ pick up different dynamical phases during their time evolution, reading-out our state at a specific time $t^*$ transforms an initially prepared product state $|eg\rangle_1 = (|0\rangle + |\pi\rangle)/\sqrt{2}$ into a desired arbitrary entangled state $\propto (\ket{eg}_1 + \alpha(t^*) \ket{ge}_1)$. In general, however, this mechanism will lead to a lower fidelity as compared with that based on the protocol described above, and as such distillation protocol might also need to be supplemented in the actual implementation of such an arbitrary entangled state generation.

	\section{Quantum state transfer}
	\label{QST}
	\begin{figure}[htb]
		\centering
		\includegraphics[angle =0,width=0.45 \textwidth]{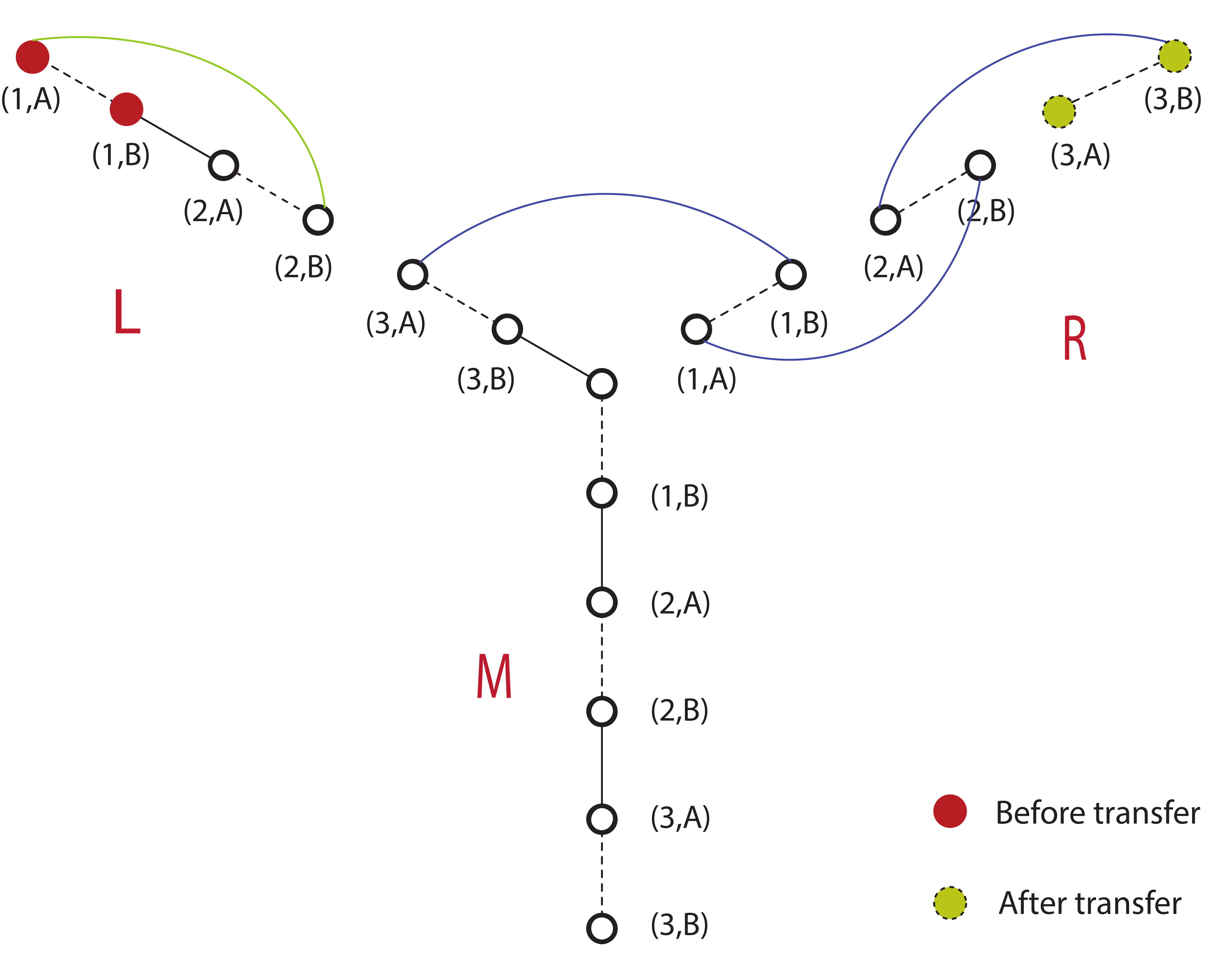}
		\caption{Schemetic diagram of the QST protocol. The green line represents the coupling added in \textit{Phase I} of QST and the blue line represents the coupling added in \textit{Phase II}. Again, dashed lines represent the coupling in $H_1$ and solid lines present the coupling in $H_2$.}
		\label{fig:QST_schematic}
	\end{figure}
	In the previous section, we have shown how entangled qubits arise as either a zero or $\pi$ mode. In this section, we propose a protocol to transfer these zero and $\pi$ modes (and thus any entangled qubits) from the leftmost end (L branch) to the rightmost end (R branch) of the chain. To this end, the importance of using Y-junction geometry to facilitate such a QST is now clear. That is, in a strictly one-dimensional chain of such Xmon qubits, both zero and $\pi$ modes necessarily emerge at its both ends. As a result, transferring a zero or $\pi$ mode from one end to the other necessarily leads to the interference with zero or $\pi$ mode located at the other end, thus destroying the transferred information. By utilizing a Y-junction geometry, we can instead adjust our system such that one pair of zero and $\pi$ modes is located at the end of L branch, whereas the other is at the end of M branch. By encoding our information in the zero and $\pi$ modes originally located in L branch, we can faithfully transfer such information to the other end of R branch, thus completing our QST procedure. Such a transfer can then be accomplished by performing a series of adiabatic manipulations, which can be divided into two phases below and summarized in Fig.~\ref{fig:QST_schematic}:
	
	\textit{Phase I}: Transferring zero and $\pi$ modes from the end of the L branch to the middle point in $(N_L - 1)/2$ steps, $N_L$ being the number of qubits in the L branch. In the $x^{\text{th}}$ step, a new term is introduced in $H_2$, i.e., $h_x^{(3)}  =j_2 \sin \phi_x\  \sigma^{(L)\dagger}_{A,2x-1} \sigma^{(L)}_{B,2x} + h.c.$, and we set the coupling strength $j_{\rm{inter},2x}=j_2 \cos\phi_x$, where $\phi_x$ is the parameter adiabatically increasing from 0 to $\pi/2$. As detailed in Appendix~\ref{ph1}, the zero and $\pi$ modes at any stroboscopic time within the $x$-th step are found as
	\begin{align}
	\ket{0}^{(L)} = \frac{1}{\sqrt{2}} \bigg[&\cos{\phi_x} \left(\ket{eg}^{(L)}_{2x-1} - \ket{ge}^{(L)}_{2x-1}\right) +\nonumber\\
	 &\sin{\phi_x} \left(\ket{eg}^{(L)}_{2x+1} - \ket{ge}^{(L)}_{2x + 1}\right) \bigg]\;,\\	
	\ket{\pi}^{(L)} = \frac{1}{\sqrt{2}} \bigg[&\cos{\phi_x} \left(\ket{eg}^{(L)}_{2x - 1} + \ket{ge}^{(L)}_{2x-1}\right) + \nonumber\\
	&\sin{\phi_x} \left(\ket{eg}^{(L)}_{2x + 1} + \ket{ge}^{(L)}_{2x + 1}\right) \bigg].	
	\end{align}
    At the end of this phase, i.e., after completing the $(N_L-1)/2$-th step, both zero and $\pi$ modes are transferred to the middle point of the Y-junction, i.e.,
	\begin{align}
	\ket{0}^{(L)}: &\ \ket{eg}_{1}^{(L)} - \ket{ge}_{1}^{(L)}\rightarrow  \ket{eg}_{N_L}^{(L)} - \ket{ge}_{N_L}^{(L)},\\
	\ket{\pi}^{(L)}: &\ \ket{eg}_{1}^{(L)} + \ket{ge}_{1}^{(L)} \rightarrow  \ket{eg}_{N_L}^{(L)} + \ket{ge}_{N_L}^{(L)}.
	\end{align}	
	\textit{Phase II}: Transferring zero and $\pi$ modes from the middle point to the end of R branch in $N_R$ steps, $N_R$ being the number of qubits in the R branch. In the $x$-th step, a new term, $h_x^{(4)}  =j_2 \sin \phi_x\  \sigma^{(R)\dagger}_{A,x-1} \sigma^{(R)}_{B,x} + h.c.$, is introduced in $H_2$, where $\phi_x$ is the adiabatic parameter swept slowly at every period from $0$ to $\pi/2$, and $\sigma^{(R)}_{A,0} \equiv \sigma^{(L)}_{A,N_L}$. As detailed in Appendix~\ref{ph2}, the zero and $\pi$ modes at any stroboscopic time within the $x$-th are
	\begin{align}
	\ket{0}^{(L)} = \frac{1}{\sqrt{2}} \bigg[&\cos(\frac{\pi}{2} \sin\phi_x) \left(\ket{eg}^{(R)}_{x} - \ket{ge}^{(R)}_{x} \right) +\nonumber\\ &\sin(\frac{\pi}{2}\sin\phi_x)(-\ket{eg}^{(R)}_{x+1} + \ket{ge}^{(R)}_{x+1}) \bigg], \\
	\ket{\pi}^{(L)} = \frac{1}{\sqrt{2}} \bigg[&\cos(\frac{\pi}{2} \sin\phi_x) \left(\ket{eg}^{(R)}_{x} + \ket{ge}^{(R)}_{x} \right) +\nonumber\\ & \sin(\frac{\pi}{2}\sin\phi_x)(\ket{eg}^{(R)}_{x+1} + \ket{ge}^{(R)}_{x+1}) \bigg].
	\end{align}
	 At the end of this phase, i.e., after completing the $N_R$-th step, the zero and $\pi$ modes are perfectly transferred to the right end of branch R,
	\begin{align}
	&\ket{0}^{(L)}:   (-1)^{N_L+N_R}(\ket{eg}_{N_R}^{(R)} - \ket{ge}_{N_R}^{(R)})\otimes G^{\prime},\\
	&\ket{\pi}^{(L)}:  (\ket{eg}_{N_R}^{(R)} + \ket{ge}_{N_R}^{(R)})\otimes G^{\prime}.
	\end{align}

	While the above protocol is presented in \emph{the ideal case}, i.e, based on special parameter values discussed in Sec.~\ref{model}, the actual implementation of our protocol does not rely on such fine tuning. Indeed, as long as the zero and $\pi$ modes in the system remain well-separated in quasienergies from the bulk states during the adiabatic manipulations (see e.g. Fig.~\ref{fig:fidelity T band}(a)), the above QST protocol is still expected to work with a good fidelity. We will discuss this aspect further in the next section.
	
	\section{Discussion}
	\label{disc}
	
	\begin{figure}[ht]
		\centering
		\includegraphics[angle =0,width=0.4 \textwidth]{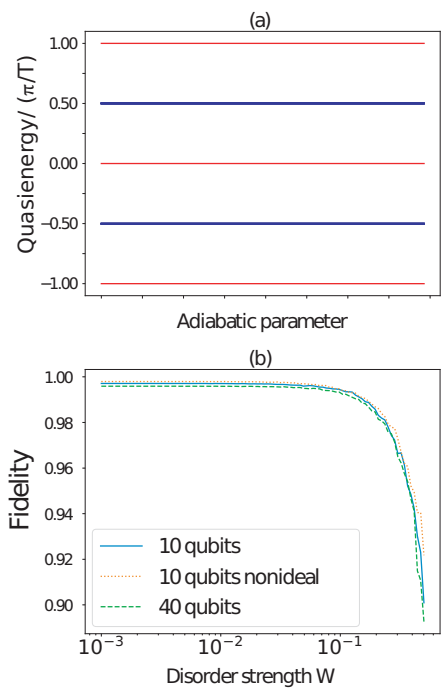}
		\caption{(a) The quasienergy spectrum of our system during QST. Notice that quasienergy zero and $\pi/T$, which correspond to our zero and $\pi$ modes respectively, remain well-separated from the other quasienergies at all times. (b) Fidelity ($|\langle \psi_i|\psi_d\rangle |$) against the disorder strength. In the 10 qubits system, we take $N_L = 6, N_R = 4, N_M = 8$, whereas in the 40 qubits system, we take $N_L = 22, N_R = 18, N_M = 8$. The parameters used for the nonideal case (the orange dotted line) are $J_1 = 1.5 i , j_2^{L} = j_2^{M} = 3 i ,j_2^{R} = -0.1 i, J_2 = j_1 = 0$. Each step takes 40 (70) periods to complete in the 10 (40) qubits system and each data point is averaged over $100$ disorder realizations.}
		\label{fig:fidelity T band}
	\end{figure}

	In practice, perfect modulation of the coupling strengths is impossible. As such, we will now examine the robustness of the QST protocol presented in Sec.~\ref{QST} against coupling disorders, which are implemented by adding each of the following terms to $H_1$ and $H_2$ respectively,
	\begin{align}
	\Delta H_1 &= \sum_{c\in C}\sum_{m} \left( \delta_{1,m} J_1~\sigma^{c\dagger}_{B,m}\sigma^{c}_{A,m} +h.c.\right),\nonumber\\
	\Delta H_2 &= \bigg[\sum_{c\in \{L,M\}}\sum_{m} ( \delta_{2,m} j_2~\sigma^{c\dagger}_{B,m}\sigma^{c}_{A,m+1} + h.c.)  + \nonumber
	\\& \sum_{m^{\prime}}(\delta_{3,m} j_2 \sigma^{L\dagger}_{A,2m^{\prime}-1} \sigma^{L}_{B,2m^{\prime}} + h.c.)+  \nonumber\\ &\sum_{m^{\prime \prime}} (\delta_{4,m} j_2 \sigma^{R\dagger}_{A,m^{\prime\prime}-1} \sigma^{R}_{B,m^{\prime\prime}} + h.c. )\bigg],
	\end{align}
	where $\delta_{i,m}$ is a uniform random number taken $\in [-0.5W,0.5W]$ and $W$ is the disorder strength. In addition to disorders introduced in the original system,  we further consider the presence of disorders during the numerical implementation of QST protocol introduced in Sec.~\ref{QST}. This is accomplished by modifying the newly added couplings $h_x^{(3)}  =\sin \phi_x j_2(1 + \delta_x^{(3)})  \sigma^{(R)\dagger}_{A,2x-1} \sigma^{(R)}_{B,2x} + h.c.$ and $h_x^{(4)}  =\sin \phi_x j_2(1+\delta_x^{(4)})  \sigma^{(R)\dagger}_{A,2x-2} \sigma^{(R)}_{B,2x - 1} + h.c.$.
	
	By denoting the transferred state as $\ket{\psi_i}$ in the ideal case and $\ket{\psi_d}$ in the case with disorders, we numerically calculate fidelity $F = \abs{\braket{\psi_i}{\psi_d}}$ as a function of the disorder strength in Fig.~\ref{fig:fidelity T band}(b). In addition to the robustness of our QST protocol against small to moderate disorders, the orange line of Fig.~\ref{fig:fidelity T band}(b) also demonstrates the good performance of our QST protocol at other system parameters, such as $J_1 = 1.5 i , j_2^{L} = j_2^{M} = 3 i ,j_2^{R} = -0.1 i, J_2 = j_1 = 0$, which deviate rather significantly from \emph{the ideal case} in which our QST protocol is analytically solvable.
	\\

	\begin{figure}[htb]
		\centering
		\includegraphics[angle =0,width= 0.4 \textwidth]{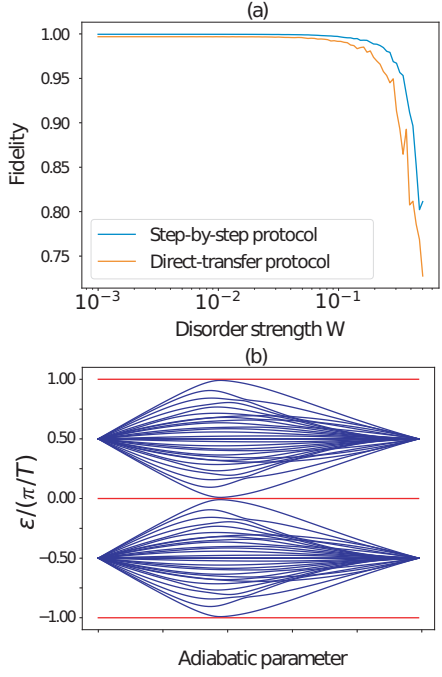}
		\caption{(a) Comparison of the fidelity against disorder strength between our proposed step-by-step protocol in Sec.~\ref{QST} and the direct-transfer protocol. Both system has the size $N_L = 30, N_R = 30, N_M = 8$ and the total time for both QST is 6600 periods. (b) Quasienergy spectrum of the direct-transfer protocol during the QST process. Notice that the quasienergy gap vanishes somewhere during the process, leading to the corruption of the transferred information and thus lower fidelity. Quasienergy spectrum of the step-by-step protocol can be referred to FIG. \ref{fig:fidelity T band} (a).}
		\label{fig:compare}
	\end{figure}
	
	We also note that Phase I and Phase II of our QST protocol can in principle be sped up by performing the actions in all $(N_L-1)/2$ and $N_R$ steps, respectively, in one go. That is, starting with $|0\rangle^{(L)}$ and $|\pi\rangle^{(L)}$ localized at the left end of the L branch, one can introduce $h^{(3)}=\sum_x h_x^{(3)}$ in $H_2$ and take $j_{\rm inter, 2}=j_{\rm inter, 4}=\cdots = j_{\rm inter, N_L-1}=j_2 \cos\phi$ simultaneously to move $|0\rangle^{(L)}$ and $|\pi\rangle^{(L)}$ to the middle point in Phase I, followed directly with the introduction of $h^{(4)}=\sum_x h_x^{(4)}$ in $H_2$ to further send $|0\rangle^{(L)}$ and $|\pi\rangle^{(L)}$ to the right end of R branch in Phase II. While this approach works very well for sufficiently small systems, increasing the number of qubits will inevitably cause the transferred state to become more delocalized in the middle of such a direct-transfer protocol, leading to the unavoidable closing of the quasienergy spectrum as shown in Fig.~\ref{fig:compare}(b). By contrast, the step-by-step QST protocol introduced in Sec.~\ref{QST} ensures that the transferred state remains localized at all times, thus maintaining large quasienergy gaps between zero or $\pi$ modes and the bulk quasienergies, even at a very large number of qubits. In such cases, the step-by-step protocol is expected to perform better as compared with the direct-transfer protocol, which we have also verified in Fig.~\ref{fig:compare}(a) for the case of $68$ qubits.
	

	\begin{figure}[ht]
		\centering
		\includegraphics[angle =0,width=0.4\textwidth]{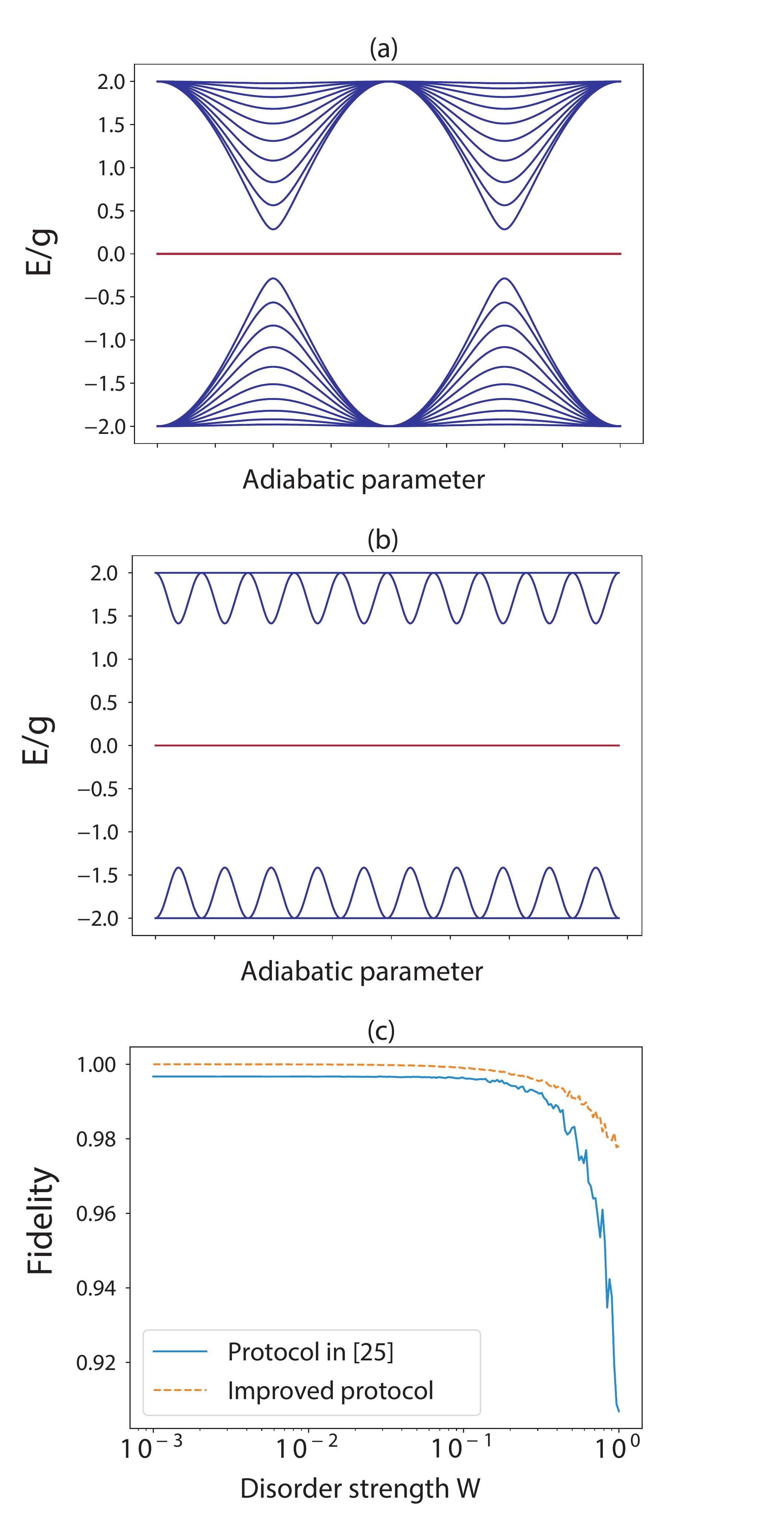}
		\caption{(a) The energy band spectrum of a single qubit transfer along a chain of 21 Xmon qubits of Ref.~\cite{tqs3} by performing the transfer in one go. (b) Same as in panel (a), but we divide the QST process in $20$ steps; (c) Comparison of the fidelity as a function of the disorder strength between the two protocols in panels (a) and (b). The total adiabatic time for both QST protocols is $t_{\rm tot} = \pi/(0.01g).$ }
		\label{fig:Comparison_ZSL}
	\end{figure}
	Inspired from the above analysis, we may also propose an improvement to the QST protocol introduced in \cite{tqs3}. In particular, Ref.~\cite{tqs3} proposes a similar QST protocol by using a chain of Xmon qubits with time-independent coupling. Due to the lack of $\pi$ modes in static systems, however, the use of dimerized coupling in such a chain only enables the transfer of a single qubit from one end to the other, which Ref.~\cite{tqs3} proposed to accomplish in one step by simultaneously modulating all the qubit-qubit couplings. While their results show a good QST fidelity at small number of qubits, the same problem of vanishing energy gap will also arise at larger number of qubits. As such, the idea of breaking down QST process into steps in the spirit of our protocol in Sec.~\ref{QST} can also be adapted to enable high fidelity transfer of one qubit in such a static system scenario. To this end, we recall the static Hamiltonian used in Ref.~\cite{tqs3} describing the dimerized qubit-qubit couplings in a chain of Xmon qubits,
	
	\begin{align}
	\hat{H} = \sum_{j = 1}^{N}\left(J_0^{j}\hat{\sigma}_{A,j}^{\dagger}\hat{\sigma}_{B,j} + J_1^{j}\hat{\sigma}_{B,j}^{\dagger}\hat{\sigma}_{A,j+1}+ h.c.      \right). \label{ZSL}
	\end{align}
	In Ref.~\cite{tqs3}, QST is accomplished by adiabatically tuning all $J_1^j$ and $J_0^j$ simultaneously as $J_i^{j} = g\left( 1 + (-1)^{i} \cos \theta \right)$, with $\theta$ being the adiabatic parameter swept from $0$ to $\pi$. In our proposed improvement, we may instead break down the QST protocol into $(N-1)$ steps. In the $x$-th step, we take $J_i^{x} = g\left( 1 + (-1)^{i} \cos \theta_x \right)$ while keeping the other coupling strengths constant, with $\theta_{x}$ being the same adiabatic parameter swept from $0$ to $\pi$. This amounts to transferring a qubit from the $x$-th unit cell to the $(x+1)$-th unit cell, so that after the $(N-1)$-th step, the qubit originally at the left end of the lattice is perfectly transferred to the right end. In this improved QST protocol, the gap in the energy spectrum is significantly larger than the original protocol, as illustrated in Fig.~\ref{fig:Comparison_ZSL}(a) and (b). To further test the robustness of this improved protocol, we again consider the presence of disorders by adding the following term to the original Hamiltonian of Eq.~\ref{ZSL},
	\begin{align}
	\hat{H}_{d} = \sum_{j} \left(J_{0}^{j} \delta_{0,j} \hat{\sigma}_{A,j}^{\dagger}\hat{\sigma}_{B,j} + J_{1}^{j} \delta_{1,j} \hat{\sigma}_{B,j}^{\dagger}\hat{\sigma}_{A,j+1} + h.c. \right),
	\end{align}
	where $\delta_{0,j}$ and $\delta_{1,j}$ are random numbers taken $\in [-0.5W,0.5W]$ and $W$ is the disorder strength. The fidelity as a function of the disorder strength is plotted in Fig \ref{fig:Comparison_ZSL}(C). To have a fair comparison, the total transfer time is the same in for the two protocols, $t_{\rm tot} = \pi/(0.01g)$. It is clear that our proposed protocol indeed improves the robustness of such a system during QST. To conclude, breaking down QST process into steps is one of our main results in this paper, which can be applied to improve the fidelity of adiabatic-based QST for large system sizes, both in time-periodic and static settings. This in turns enables us, at least in principle, to transfer qubits over an arbitrarily large distance.

	\section{Concluding Remarks}
	\label{conc}
	
	In this paper, we have proposed an innovative scheme to realize high-fidelity and long-distance transfer of an entangled state along a Y-shaped topologically non-trivial qubit chain in the presence of periodic driving. Before the state is transferred, a maximal entangled state is prepared through an adiabatic process, in which a key step is to introduce a nontrivial rotation between zero and $\pi$ modes (both being topological edge states of the qubit chain).  In the ideal situation, our QST can perfectly transfer an entangled state from one branch to another branch. In a more realistic situation, where disorder effects are introduced, the transfer fidelity is found to be robust against random noise, due to the inherent robustness of encoding qubits built from Floquet zero and $\pi$ edge modes.  Furthermore, one important property of our QST scheme is that the gap between the involved zero and $\pi$ modes and the bulk states in the quasienergy spectrum does not scale down to zero as the size of the qubit chain increases.   Thus, our scheme enables us to transfer the entangled qubits over long distance without the loss of topological protection or adiabaticity. Inspired by our QST scheme, we have also improved the QST protocol proposed in \cite{tqs3}. Indeed, one simple modification over the original protocol greatly enhances its robustness against disorder and also makes it possible to realize long-distance QST, but for single-qubit states only.

 The potential applications of our QST protocol should lie in solid-state based quantum information processing and quantum computation, where entangled qubits need to be transferred in certain solid state devices over a not-necessarily short distance.  Given that topological edge modes, especially those of periodically driven systems, are already found to have great potential in implementing quantum computation protocols \cite{RG,RG2,RG6}, it is stimulating to see by now that Floquet topological edge modes can further facilitate entangled state transfer along solid-state-based qubit chains.

	\vspace{0.3cm}

	\n {\bf Acknowledgements:} J.G. acknowledges fund support by the Singapore NRF
Grant No. NRF-NRFI2017- 04 (WBS No. R-144-000-378-
281) and by the Singapore Ministry of Education Academic
Research Fund Tier-3 (Grant No. MOE2017-T3-1-001 and
WBS. No. R-144-000-425-592).

	\appendix
	
\begin{widetext}
	
\begin{center} {\bf Appendix}  \end{center}
 This Appendix consists of two parts. In Appendix A,  we analytically compute the two topological invariants $\upsilon_0$ and $\upsilon_\pi$ in our system, which determine the number of zero and $\pi$ modes. In Appendix B,  we present the derivation of the zero and $\pi$ modes at any stroboscopic time during the transfer protocol in Sec.~\ref{QST}.

	\section{Analytical calculation of $\upsilon_{0}$ and $\upsilon_{\pi}$}
	\label{zpical}
	
	We calculate the topological invariants $\upsilon_{0}$ and $\upsilon_{\pi}$ of our system (see Eq.~\ref{ham2}) under the condition $J_2 = j_1 = 0$. As we discuss in the main text, $\upsilon_{0}$ can be expressed in terms of $b(k)$ in the $F$ matrix, which is explicitly given by
	\begin{align}
	b(k) = e^{i k }\sin(\frac{j_2}{2}) \cos(\frac{J_1}{2}) + \cos(\frac{j_2}{2}) \sin(\frac{J_1}{2}).
	\end{align}
	Therefore,
	\begin{align}
	\upsilon_{0} &= \frac{1}{2\pi i}\int_{-\pi}^{\pi} dk\ b^{-1}\frac{d}{dk} b\\
	& = \frac{1}{2\pi}\int_{-\pi}^{\pi} dk \ \frac{1}{1 + \frac{\tan{J_1/2}}{\tan{j_2/2}} e^{-i k}}
	\end{align}
	It can be shown that for any real number $A$,
	\begin{align}
	\frac{1}{2\pi}\int_{-\pi}^{\pi} dk \frac{1}{1 + A e^{-i k}}  =
	 \begin{cases}
	1 & \abs{A} < 1 ,\\
	0 & \abs{A} > 1.
	\end{cases}
	\end{align}
	Since $\tan{x}$ is monotonically increasing in the range $x \in (0,\pi/2)$, the number of edge states with zero quasienergy (zero modes) is
	\begin{align}
	\upsilon_{0} =
		\begin{cases}
		1 & j_2 > J_1 ,\\
		0 & j_2 < J_1.
		\end{cases}
	\end{align}
	The topological invariant $\upsilon_{\pi}$ can be calculated in a similar manner,
	\begin{align}
	d(k) = \cos(\frac{j_2}{2}) \cos(\frac{J_1}{2}) - e^{-ik} \sin(\frac{j_2}{2}) \sin(\frac{J_1}{2}).
	\end{align}
	Therefore,
	\begin{align}
	\upsilon_{\pi} &= \frac{1}{2 \pi i} \int_{-\pi}^{\pi} dk\ d^{-1} \frac{d}{dk} d\\
	               &= \frac{1}{2 \pi}\int_{-\pi}^{\pi} dk  \frac{1}{1 - e^{ik} \cot(J_1/2) \cot(j_2/2)}
	\end{align}
	As we discussed above, $\upsilon_{\pi} = 1$ if $\abs{\cot(J_1/2) \cot(j_2/2)} < 1$. The inequality can be further simplified by using the trigonometry identity,
	\begin{align}
	\cot(J_1/2) \cot(j_2/2) = \frac{\cos(\frac{J_1 - j_2}{2}) + \cos(\frac{J_1 + j_2}{2})}{\cos(\frac{J_1 - j_2}{2}) - \cos(\frac{J_1 + j_2}{2})}.
	\end{align}
	As $\cos(\frac{J_1 - j_2}{2}) > 0$ when $0< J_1, j_2 < \pi$, we only require $j_2 + J_1 > \pi$ to have $\abs{\cot(J_1/2) \cot(j_2/2)} < 1.$ Hence, the number of edge states with $\pi/T$ quasienergy ($\pi$ modes) is
 	\begin{align}
	 	\upsilon_{\pi} =
	 	\begin{cases}
	 	1 & j_2 + J_1 > \pi ,\\
	 	0 & j_2 + J_1 < \pi.
	 	\end{cases}
 	\end{align}

	\section{Derivation of $|0\rangle^{(L)}$ and $|\pi\rangle^{(L)}$ during the QST protocol}
	\subsection{Phase I}
	\label{ph1}
	The Floquet operator $U$ of our system can be written as a product of two exponentials,
	\begin{align}
	U = \exp(-i H_2) \times \exp(-i H_1),
	\end{align}
	where the period $T$ is now set to be 2 for brevity. To simplify our notation, we will focus on the relevant terms in $H_1$ and $H_2$ which have support on the transferred zero and $\pi$ modes. In the $x$-th step, $H_1$ and $H_2$ read,
	\begin{align}
	&H_1 = i \frac{\pi}{2}\left(\sigma^{(L)\dagger}_{A,2x-1} \sigma^{(L)}_{B,2x-1}  +\sigma^{(L)}_{A,2x} \sigma^{(L) }_{B,2x} + \sigma^{(L)\dagger}_{A,2x+1} \sigma^{(L)}_{B,2x+1} + h.c. \right),\\
	&H_2 = i \pi \left(\sigma^{(L)\dagger}_{B,2x-1} \sigma^{(L)}_{A,2x} + \sin{\phi_x}\ \sigma^{(L)\dagger}_{A,2x-1} \sigma^{(L)}_{B,2x} + \cos{\phi_x}\ \sigma^{(L)\dagger}_{B,2x} \sigma^{(L)}_{A,2x+1} + h.c.\right),
	\end{align}
	where $\phi_x$ is adiabatically increased at every period from 0 to $\pi/2$. It can be verified that
	\begin{align}
    &\ket{0}^{(L)} = \frac{1}{\sqrt{2}} \left[\cos{\phi_x} \left(\ket{eg}^{(L)}_{2x-1} - \ket{ge}^{(L)}_{2x-1}\right) + \sin{\phi_x} \left(\ket{eg}^{(L)}_{2x+1} - \ket{ge}^{(L)}_{2x+1}\right) \right]\;,\\
    &\ket{\pi}^{(L)} = \frac{1}{\sqrt{2}} \left[\cos{\phi_x} \left(\ket{eg}^{(L)}_{2x-1} + \ket{ge}^{(L)}_{2x-1}\right) + \sin{\phi_x} \left(\ket{eg}^{(L)}_{2x+1} + \ket{ge}^{(L)}_{2x+1}\right) \right]\;.
	\end{align}
	Note that
	\begin{align}
	\exp(-i H_1) \ket{eg}^{(L)}_{2x \pm 1} &=  \ket{eg}^{(L)}_{2x \pm 1},\\
	\exp(-i H_1) \ket{ge}^{(L)}_{2x \pm 1} &=  \pm \ket{ge}^{(L)}_{2x \pm 1},\\
	\exp(-i H_2) \ket{ge}^{(L)}_{2x \pm 1} &= \ket{ge}^{(L)}_{2x\pm1},\\
	\exp(-i H_2) \ket{{eg}}^{(L)}_{2x\pm 1} &= \mp \cos{2\phi_x}\ket{eg}_{2x\pm 1} + \sin{2\phi_x} \ket{eg}_{2x \mp 1} .
	\end{align}
	By using Eq. (A15-19), one can directly verify that $U\ket{0} = \ket{0}$ and $U\ket{\pi} = -\ket{\pi}$.

	\subsection{Phase II}
	\label{ph2}
	In a similar fashion, we can write $H_1$ and $H_2$ in the $x$-th step as
	\begin{align}
	&H_1 = i \frac{\pi}{2}\sum_{m = 1}^{x+1} \left(\sigma^{(R) \dagger}_{A,m-1} \sigma^{(R)}_{B,m-1}  + h.c. \right) + i \frac{\pi}{2} \sum^{N_M}_{m^{\prime}=1} \left(\sigma^{(M) \dagger}_{A,m^{\prime}} \sigma^{(M)}_{B,m^{\prime}}  + h.c.\right),\\
	&H_2 = i \pi \sin{\phi_x}\ \sigma^{(R) \dagger}_{A,x-1} \sigma^{(R)}_{B,x} + i \pi \left( \sum^{x-1}_{m = 1} \sigma^{(R) \dagger}_{A,m-1} \sigma^{(R)}_{B,m} + \sum^{N_M}_{m^{\prime} = 1} \sigma^{(R) \dagger}_{A,m^{\prime}-1} \sigma^{(R)}_{B,m^{\prime}}\right) + h.c.,
	\end{align}
	where $\sigma^{(R) \dagger}_{A,0}=\sigma^{(L) \dagger}_{A,N_L}$, $\sigma^{(M) \dagger}_{B,0}= \sigma^{(R) \dagger}_{B,0} =  \sigma^{(L) \dagger}_{B,N_L}$, and $\phi_x$ is adiabatically increased at every period from 0 to $\pi/2$. From the equations
	\begin{align}
	\exp(-i H_1) \ket{eg}^{(R)}_{x (-1)} &=  \ket{eg}^{(R)}_{x (- 1)},\\
	\exp(-i H_1) \ket{ge}^{(R)}_{x (- 1)} &= - \ket{ge}^{(R)}_{x (- 1)},\\
	\exp(-i H_2) \ket{eg}^{(R)}_{x - 1} &=  \cos(\pi \sin\phi_x) \ket{eg}^{(R)}_{x - 1} + \sin(\pi\sin\phi_x) \ket{eg}^{(R)}_{x},\\
	\exp(-i H_2) \ket{ge}^{(R)}_{x-1} &=  - \ket{ge}^{(R)}_{x-1},\\
	\exp(-i H_2) \ket{eg}^{(R)}_{x} &=   \ket{eg}^{(R)}_{x},\\
	\exp(-i H_2) \ket{ge}^{(R)}_{x} &= - \sin(\pi\sin\phi_x) \ket{eg}^{(R)}_{x-1}+ \cos(\pi \sin\phi_x) \ket{ge}^{(R)}_{x},
	\end{align}it can be verified that
	\begin{align}
	&\ket{0}^{(L)} = \frac{1}{\sqrt{2}} \left[\cos(\frac{\pi}{2} \sin\phi_x) \left(\ket{eg}^{(R)}_{x-1} - \ket{ge}^{(R)}_{x-1} \right) + \sin(\frac{\pi}{2}\sin\phi_x)(-\ket{eg}^{(R)}_{x} + \ket{ge}^{(R)}_x) \right], \\
	&\ket{\pi}^{(L)} = \frac{1}{\sqrt{2}} \left[\cos(\frac{\pi}{2} \sin\phi_x) \left(\ket{eg}^{(R)}_{x-1} + \ket{ge}^{(R)}_{x-1} \right) + \sin(\frac{\pi}{2}\sin\phi_x)(\ket{eg}^{(R)}_{x} + \ket{ge}^{(R)}_x) \right].
	\end{align}

\end{widetext}

	\end{document}